\documentstyle[psfig]{mn}

\title{Premature dismissal of high-redshift elliptical galaxies}

\author[R. Jimenez et al.]
{Raul Jimenez$^{1}$, Amancio C. S. Fria\c{c}a$^{2}$, 
James S. Dunlop$^{1}$, Roberto J. Terlevich$^{3,4}$, \and 
John A. Peacock$^{1}$ \& Louisa A. Nolan$^{1}$\\
$^1$ Institute for Astronomy, The University of Edinburgh, 
Royal Observatory, Edinburgh EH9 3HJ, UK.\\
$^2$ Instituto Astron\^{o}nomico e Geof\'{\i}sico, USP, Av. Miguel Stefano
 4200, 04301-904 S\^{a}o Paulo, SP, Brasil.\\
$^3$ Royal Greenwich Observatory, Madingley Road, Cambridge CB3 0EZ, UK.\\
$^4$ Visiting Professor at Instituto Nacional de Astrof\'{\i}sica, 
Optica y Electr\'{o}nica. Av. Luis Enrique Erro 1, Tonanzintla, Puebla, 
Mexico.}

\begin{document}
\maketitle 

\begin{abstract}
It has recently been argued that single-collapse high-redshift models for
elliptical galaxy formation can be rejected because they predict large numbers
of very red galaxies at intermediate redshifts which are not seen in deep
optical-infrared surveys. We argue, however, that this conclusion is premature
since, while much effort has been invested in refining the predictions of
hierarchical CDM models, only very simplistic models have been used to study
the evolution of galaxies in other cosmogonies (e.g. isocurvature models). We
demonstrate that the use of a more realistic multi-zone chemo-dynamical
single-collapse model, yields colours at intermediate redshifts which are much
bluer than inferred from the one-zone model, and indeed are comparable to
those predicted by hierarchical merging despite still allowing $> 90\%$ of the
final stellar mass of elliptical galaxies to be formed in the first Gyr of
their evolution. We, therefore, conclude that the one-zone model should be
avoided to predict the colours of high-redshift galaxies and that the use of
realistic multi-zone models allows the existence of ellipticals at high
redshift, being their dismissal premature.
\end{abstract}

\section{Introduction}
The epoch of formation of elliptical galaxies remains a fundamental and
controversial issue in cosmology.  While observations show that at least some
galaxies were already in place at redshifts $ z \simeq 5$ (Dunlop et al. 1996;
Franx et al. 1997; Dey et al. 1998; Hu et al. 1998), other recent data (Zepf
1997; Abraham et al. 1998; Glazebrook et al. 1998; Kauffman \& Charlot 1998)
have been interpreted as indicative of a general formation epoch at $ z <
3$. In hierarchical models of galaxy formation dominated by cold dark matter
(CDM), elliptical galaxies arise from the merging at low redshift of
intermediate mass disks (e.g. Kauffmann, Charlot \& White 1996; Baugh, Cole \&
Frenk 1996).  In contrast, in isocurvature CDM (Peebles 1998) (ICDM), or warm
dark matter (Colombi, Dodelson \& Widrow 1995) (WDM) models, ellipticals are
assembled at $ z > 3$ in a major single collapse event. Zepf (1997) and Barger
et al. 1998 have recently argued that such models can be rejected because they
predict large numbers of very red galaxies at intermediate redshifts which are
not seen in deep optical-infrared surveys. However, this conclusion maybe
premature since, while detailed predictions (e.g. Baugh, Cole \& Frenk 1996)
have been made for hierarchical CDM models, only very simplistic models have
been used to study the evolution of galaxies in ICDM or WDM models.  In
particular the physically unrealistic `one-zone' model (Baade 1957; Arimoto \&
Yoshii 1987; Matteucci \& Tornambe 1987) (or monolithic model), in which
elliptical galaxies are formed in a single $ 10^8$-yr starburst predicts very
red colours at all intermediate redshifts (Zepf 1997; Barger et al. 1998). The
aim of this paper is to explore a more realistic multi-zone chemo-dynamical
single-collapse model for predicting the colours of high-redshift elliptical
galaxies.

\section{The model}
 
We have used the chemo-dynamical code developed by Fria\c{c}a \& Terlevich
(1998) to compute a single collapse multi-zone model in which a single massive
dark halo hosts baryonic gas that will fall into the center of the dark
potential and will subsequently form stars. The code assumes spherical
symmetry and thus uses a 1D hydro-dynamical solver that calculates the
hydro-dynamical evolution and radial behaviour of the ISM and stars during the
early collapse and evolution of the elliptical galaxy. The galaxy is divided
into 100 zones from the core up to the tidal radius. We assume that the
specific star formation rate (SF) follows a power-law function of gas density
($\rho$); ${\rm SF}(r,t)=\nu_0 (\rho/\rho_0)^{1/2}$ (here $\nu_0=10$
Gyr$^{-1}$, $\rho_0$ is the initial average gas density inside the core radius
of the dark halo, and $r$ and $t$ are the spatial and time coordinates
respectively). We have also included inhibition of star formation by
inefficient gas cooling and expanding gas. A characteristic feature of these
models is that several episodes of inflow and outflow occur simultaneously at
different radii (Fria\c{c}a \& Terlevich 1998). The detailed behaviour of
these models of course depends on the precise parameters chosen. Although we
will show here results for specific cases, we believe that the critical
feature is robust: the global star formation history in the galaxy naturally
extends over a period of about 1 Gyr.

\begin{figure}
\centerline{
\psfig{figure=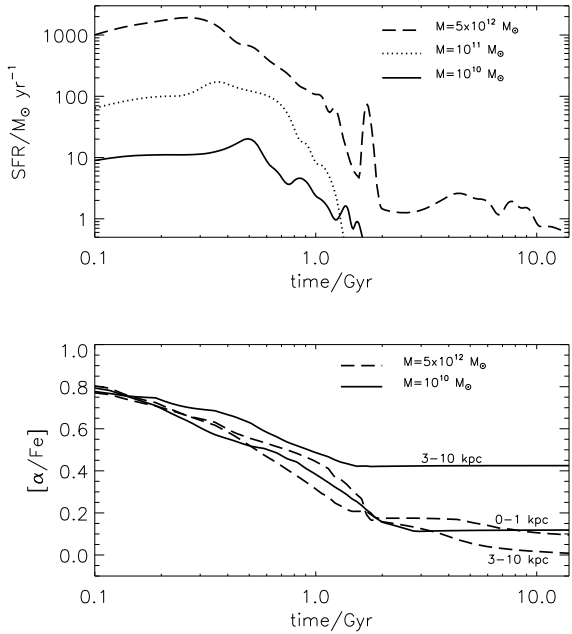,height=9cm,angle=0}}
\caption{The top panel shows the SF rate inside 10 kpc for our multi-zone
model for three different masses. The primary epoch of star formation lasts
for about 1 Gyr in each case, but, particularly in the most massive galaxy,
low-level star-formation continues until the present epoch.  Contrary to what
is often claimed in the literature, a moderate extended period of star
formation is consistent with the observed $[\alpha/Fe]$ ratios in elliptical
galaxies. Since most of the stellar population is formed in a short burst of
duration $\sim 0.5$ Gyr, the ISM is not contaminated by SNIa. The bottom panel
illustrates this for two different masses, where the time evolution (not age)
since the formation of the elliptical (compare with top panel) is shown. Note
that 90\% of the stars are formed at times smaller than 2 Gyr, thus the galaxy
is [$\alpha$/Fe] enhanced (about 0.4). Only stars born (less than 10\%) after
2-3 Gyr have moderate [$\alpha$/Fe] ratios, but these do not contribute to the
overall galaxy abundance.}
\end{figure}

Using the resulting distribution of the gas flow, the equations of chemical
evolution are then solved. Once the structure of the ISM and the chemical
abundances in each of the 100 zones have been computed, we use the
spectral-synthesis code developed by Jimenez et al. (1998; 1999) to compute
the spectral energy distributions of the multi-zone model at different radii
and ages. Our models constitute a realistic family which successfully account
for several observational properties of elliptical galaxies (Jimenez et
al. 1999, in preparation): (1) the observed metallicity gradients are
naturally reproduced --- the core has super-solar metallicity and the outer
regions have sub-solar metallicities; (2) the [$\alpha$/Fe] ratio is
supersolar in the core of the galaxy, and also when averaged over the whole
galaxy; (3) the observed positive mass-metallicity relation is reproduced; (4)
galactic winds from elliptical galaxies can explain the amount of iron in the
intra-cluster medium (ICM) of rich clusters of galaxies.  The time scales for
chemical enrichment are also consistent with the observations. For example,
the high-metallicity core is formed quickly enough to explain the strong metal
lines observed in $z\sim 5$ QSOs (assuming that a young elliptical core hosts
the QSO event), and the ICM is enriched rapidly enough to account for the lack
of evolution of the ICM Fe abundance in rich galaxy clusters (Mushotzky \&
Loewenstein 1997) up to $z \sim 0.3$.  In addition, our multi-zone model does
not display the extremely high luminosities ($L_{\rm bol}$ up to $10^{15}$
L$_{\odot}$) of the one-zone models, which have never been observed
(Fria\c{c}a \& Terlevich 1998). Furthermore the model reproduces the observed
spectral ages of red elliptical galaxies found at intermediate redshifts
(Dunlop et al. 1998).

\section{Results}

One of the firm predictions of our model is that star formation in elliptical
galaxies lasts longer than $10^8$ years. In the top panel of Fig.~1 we show
the star formation rate in $M_{\odot}$ ${\rm yr}^{-1}$ for three different
galaxy masses. In general our elliptical galaxy model exhibits three
star-formation stages: stage (1) during which the bulk of the stellar
population is formed (half of the present day stellar population inside 10 kpc
is formed in 0.42, 0.40 and 0.30 Gyr for the galaxies with $10^{10}$,
$10^{11}$, and $5\times 10^{12}$ M$_{\odot}$, respectively); the duration of
this stage is well traced by the early, major peak, in the evolution of the
star formation rate (see Fig.~1 top panel); stage (2), which corresponds to a
extended period of star formation in the core (2-3 Gyr) as a result of the
inner cooling flow. This episode has more modest star-formation
rates than stage (1) and explains the blue colours of our model and thus the
absence of passively evolving elliptical galaxies in deep optical and
near-infrared surveys as claimed by Zepf (1997) and . Finally, after stage (2)
is finished, our model evolves {\it passively}.  However, very massive
galaxies retain the inner (like e.g. cD galaxies in the center of clusters)
cooling flow until the present epoch, which results in the occurrence of
subsequent low-level bursts of star formation. For the intermediate mass
galaxy the SF peaks at about 100 $M_{\odot}$ ${\rm yr}^{-1}$. On the other
hand, the most massive galaxies exhibit much higher maximum star-formation
rates, typically several hundred ${\rm M_{\odot} yr^{-1}}$. Such a maximum
star-formation rate is an order of magnitude smaller than that predicted by
the one-zone model ($\simeq 10000\, {\rm M_{\odot} yr^{-1}}$), and is
comparable to the maximum SF rates inferred for massive galaxies at $z \simeq
4$ from sub-millimetre observations (Dunlop et al. 1994).

\begin{figure*}
\centerline{
\psfig{figure=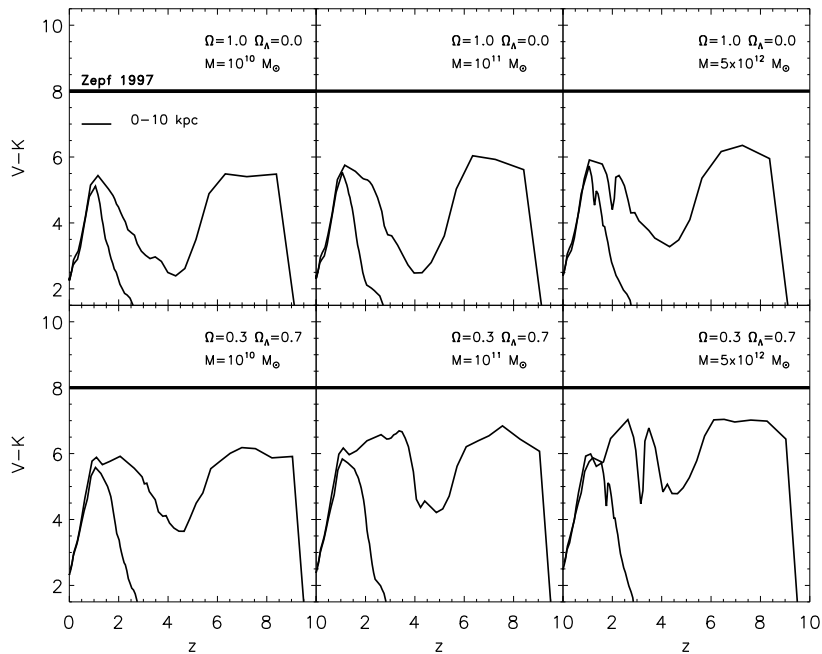,height=10cm,angle=0}}
\caption{The time-evolution of $V-K$ for three different masses and two
cosmologies. We have plotted the time-evolution for the mass fraction
contained inside a radius of 10 kpc. We have also assumed two formation
redshifts: 3 (left line) and 10 (right line) for each mass. It is worth noting
that in none of the cases the predicted $V-K$ colours for the mass inside 10
kpc (the most realistic case, since at high redshift most of the observed
light is inside 10 kpc), are redder than $V-K \approx 7$, two magnitudes
bluer! than the expectations from the monolithic model as used in Zepf
(1997). This strongly suggest that the dismissal of high redshift elliptical
galaxies is premature.}
\end{figure*}

A recognized strength of the one-zone model is that a relatively short-lived
burst of star formation seems to be required to avoid substantial
contamination of the stellar population by Fe ejected into the ISM by
supernovae SNIa (most of which explode after several $10^8$ years), which
makes it hard to reproduce the observed super-solar [$\alpha$/Fe]
abundances (where the $\alpha$ elements are: O, Ne, Mg, Si, S, Ar and
Ca) in elliptical galaxies.  Conversely, a problem in the hierarchical picture
is that, due to the extended period over which significant star formation
occurs, the resulting models of elliptical galaxies struggle to exhibit
super-solar $[\alpha$/Fe$]$ ratios, but rather tend towards the solar ratio
due to SNIa contamination of the ISM.  Since our multi-zone model has a
somewhat more extended period of star formation (Fig.~1 top panel) than the
one-zone model, it might be expected that contamination by SNIa would also
prevent the multi-zone models from reproducing the observed $[\alpha$/Fe$]$
ratios.  In fact this is not the case. In Fig.~1 (bottom panel) we show the
time-evolution of $[\alpha$/Fe$]$ for two different masses and two radii. As
expected, the predicted $[\alpha/Fe]$ ratios are initially very high due to
the fact that SN II's dominate the chemical enrichment in the early evolution
of the galaxy. Then, as later SNI explosions enrich the ISM with Fe, the
average $[\alpha/\rm Fe]$ of formed stars decreases, but the star formation is
actually switched off early enough to fix the $[\alpha/\rm Fe]$ at super-solar
values and it remains so until the present epoch.  The reason for this is that
the SNIa rate peaks during the late stage of continuing star formation (the
maximum of the SNIa rate occurs at 0.85, 0.75, and 1.8 Gyr for the models with
$10^{10}$, $10^{11}$, and $5\times 10^{12}$ M$_{\odot}$, respectively), and by
that time most of the stellar mass of the galaxy has been formed
($>90\%-95\%$). Therefore, the resulting galaxies still exhibit super-solar
$[\alpha$/Fe$]$ ratios.

The consequence of the relatively extended SF history in this model is that
elliptical galaxies do not become extremely red. Fig.~2 shows the redshift
evolution of $V-K$ for three different galaxy masses and two alternative
cosmological scenarios: an Einstein-de Sitter Universe ($\Omega_m=1.0$,
$\Omega_{\Lambda}=0.0$) and a vacuum-energy dominated Universe
($\Omega_m=0.3$, $\Omega_{\Lambda}=0.7$). These two example cosmologies have
been chosen because they represent extreme cases of a young and an old
Universe.  In both cases we have adopted a value for the Hubble constant of 65
km s$^{-1}$ Mpc$^{-1}$. In what follows we assume that all dark haloes
collapse at a given redshift and then investigate whether, as the resulting
galaxies evolve, they ever display observed optical-infrared colours redder
than the limit considered by Zepf (1997) for the monolithic model, which
translates into $V-K < 8$. For simplicity two formation redshifts are
considered: $z=10$ and $z=3$. Note that at $z>3$ the observed colours of
galaxies determined by aperture photometry are usually roughly equivalent to
their integrated colours, i.e. including the whole visible radius.

\begin{figure}
\centerline{
\psfig{figure=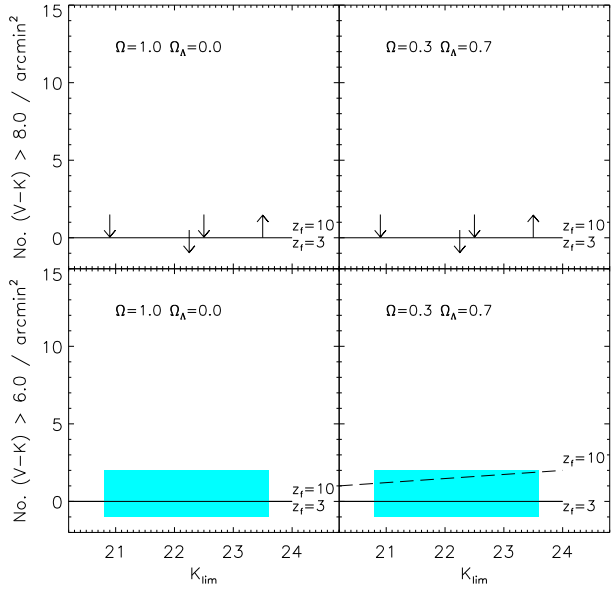,height=9cm,angle=0}}
\caption{The abundance of faint galaxies in the field, collected by
Zepf(1997), is compared with the predictions of the multi-zone model for
formation redshifts of 10 (dashed line) and 3 (continuous line). For a colour
cut-off of $V-K > 8$ the multi-zone model predicts {\it zero} galaxies (even
for $z_f=10$) in superb agreement with the observations. Moreover, even at a
much lower colour cut-off ($V-K > 6$), the multi-zone model predicts a very low
surface density of red objects, in agreement with the observations (now
displayed as a shadow region).}
\end{figure}

Fig.~2 shows that, even assuming a formation redshift of $z = 10$, the
multi-zone model never produces colours as red as $V-K = 8$. Indeed it is only
for the most massive galaxy model that a colour as red as $V-K = 7$ is ever
achieved, and that requires the assistance of a very high formation redshift
in an old (in this example $\Lambda$-dominated) universe. Thus, integrated
within a radius of 10 kpc, the maximum $V-K$ colour predicted by our
multi-zone models is typically $V-K \simeq 6.5$, comparable with the observed
colours of the reddest known ellipticals at $ z > 1$.  Of course the maximum
predicted colour is still bluer if a lower redshift of formation is assumed,
and if $z_f = 3$, the maximum predicted colour peaks at $V-K \simeq 6$ between
$z = 1 $ and $z = 2$.  It is thus clear that even if {\it all} elliptical
galaxies were to start forming stars at $z=10$, they would never become as red
as the one-zone model would predict. There is one and only reason for this:
the extended period of star formation in the multi-zone model.  Note that
these blue colours are {\it independent} of the suite of spectrophotometric
models chosen to compute them, since the effect of choosing different codes
will be at the most 0.3 magnitudes (see e.g. Jimenez et al. 1999), totally
irrelevant for the present discussion. In other words, small differences in
rest frame colours have no effect in observed colours. A direct comparison
with the observed data collected by Zepf (1997) on the abundance of faint red
galaxies is shown in Fig.~3. The top two panels show that the multi-zone model
predicts {\it zero} galaxies with $V-K > 8$, in superb agreement with the
observations. An even more robust prediction of the multi-zone model is that
the space density of galaxies with $V-K > 6$ (two magnitudes below the colour
cut-off chosen by Zepf(1997)) will be smaller than five galaxies per
arcmin$^2$ again in superb agreement with the data.

\begin{figure}
\centerline{
\psfig{figure=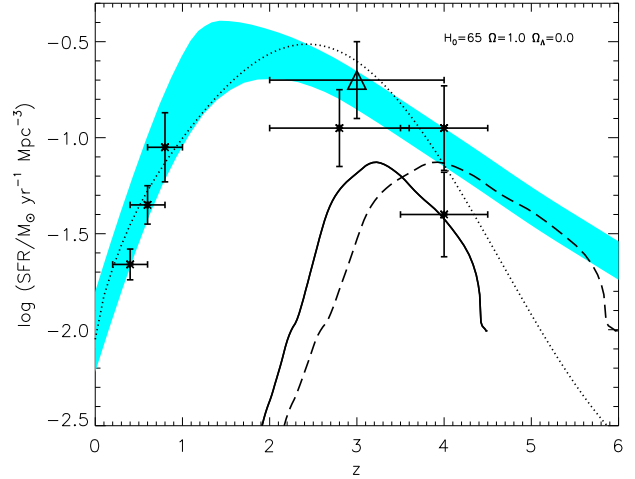,height=7cm,angle=0}}
\caption{The figure displays the global SF in the Universe assuming that all
elliptical galaxies form at a certain redshift: 4.5 (solid line) and 6 (dashed
line).  Also plotted are several observational indicators of the cosmological
evolution of star-formation density, derived from Lyman-limit galaxies in the
HDF(asterisks, the point at $z=4$ has been also plotted (highest point)
including the correction from incompleteness due to colour selection in the
HDF (C. Steidel, private comm.)), from recent deep sub-millimetre
surveys(triangle), from the cosmological evolution of luminosity density from
radio-sources(dotted line), and from the chemical evolution of metals in the
Universe (shaded region). It transpires from the plot that, even if all
elliptical galaxies were formed at redshift 6, their predicted contribution to
global star-formation density would still be consistent with the existing data
and the other external indicators and there would even be room for the bulges
of spiral galaxies to be formed at high redshift.}
\end{figure}

Finally, we consider the star-formation rates estimated from the single
collapse multi-zone model in the context of the cosmological evolution of
global star formation rate in the Universe. Several authors have recently
attempted to determine the global level of star formation in the Universe at
redshifts higher than two (Pei \& Fall 1995; Dunlop 1998; Pettini et al. 1998;
Hughes et al. 1998); the main uncertainty in this case is the precise impact
of dust at high redshift. To predict the star-formation density contributed at
high redshift by young ellipticals from the single collapse multi-zone model,
we have assumed that the number density of massive ellipticals at
high--redshift is the same as at the present--day (Terlevich \& Boyle 1993),
and that they all collapse at the same redshift. Therefore we adopt a
Schechter luminosity function given by
$\Phi(M_B)dM_B=\Phi^*10^{0.4(M^*_B-M_B)^{\alpha}}\exp[-10^{0.4(M^*_B-M_B)}]$
with $\alpha=0.23$, $M^*_B=-21$ and $\Phi^*=2.9 \times 10^{-3} h^3 Mpc^{-3}$
with the same power law extension for $M_B \le -22.5$ adopted in Fria\c{c}a \&
Terlevich (1998). Integrating this luminosity function for galaxies brighter
than $0.1 L_*$, it is possible to predict the global star-formation density of
the Universe at any redshift using our multi-zone models for different galaxy
masses. Fig.~4 shows, for an Einstein-de Sitter Cosmology, and two different
formation redshifts ($z_f = 4.5$; solid curve, and $z_f = 6$; dashed curve),
the predicted contribution of young ellipticals to the global star-formation
density at high--redshift. For comparison, we also show several observational
indicators of the cosmological evolution of star-formation density, derived
from Lyman-limit galaxies in the HDF (Pettini et al. 1998), from recent deep
sub-millimetre surveys (Hughes et al. 1998), from the cosmological evolution
of luminosity density from radio--sources (Dunlop 1998), and from the chemical
evolution of metals in the Universe (Pei \& Fall 1995).  Fig.~4 demonstrates
that our models do not conflict with the existing data, and other external
indicators of star-formation density at high redshift. Even galaxy formation
redshifts as high as 6 appear perfectly consistent with the available
data. Note that we have used an extreme case in which {\it all} galaxies form
at the same redshift. If elliptical galaxies form over a redshift range
between 3 and 10 with a small fraction of them being formed at redshift higher
than 6, then there would be even room to form bulges of spirals at a high
redshift.

\section{Conclusions}

We conclude by re-emphasizing that the hierarchical picture and the multi-zone
model considered here both predict a complete absence of objects with colours
as red as $V-K \simeq 7$ in deep optical and near-infrared surveys (unless, of
course, the intrinsic colours of the galaxy in question are heavily distorted
by dust).  Furthermore the multi-zone model does display some late episodes of
low-level star formation that could account for the observed properties of
some elliptical galaxies (Abraham et al. 1998; Glazebrook et al. 1998) in the
HDF which appear to show evidence of some recent star-forming activity and it
does not seem implausible that field and cluster ellipticals may follow a
different evolutionary path. However, in the multi-zone model considered here,
$> 90$\% of the final stellar population of the final elliptical galaxy is
formed during the first Gyr of its evolution. This provides a natural
explanation for the homogeneous properties displayed by present-day elliptical
galaxies, without violating the colour limits derived by Zepf (1997) at
intermediate redshifts.  The lack of very red galaxies at high redshifts is
therefore not a strong test of hierarchical structure formation.  Better
probes of such theories come either from the homogeneity of the
colour--magnitude relation (Bower, Kodama \& Terlevich 1998), or from a direct
counting of luminous galaxies at high redshifts. At present, there is some
evidence that the galaxy population was not created in its entirety at very
high redshifts (Kauffmann \& Charlot 1998), but single-collapse models for the
spheroidal population can not yet be ruled out (e.g. Totani \& Yoshii
1998). Larger complete samples with HST morphologies will be needed before we
can claim to have observed merging at work in modifying the galaxy population.

\end{document}